# Mechanics and Structural Stability of the Collagen Triple Helix

Michael W.H. Kirkness,[1] Kathrin Lehmann[2] and Nancy R. Forde[1,2]

[1]Department of Molecular Biology and Biochemistry and [2]Department of Physics, Simon Fraser University. Burnaby, BC, V5A 1S6 Canada

nforde@sfu.ca

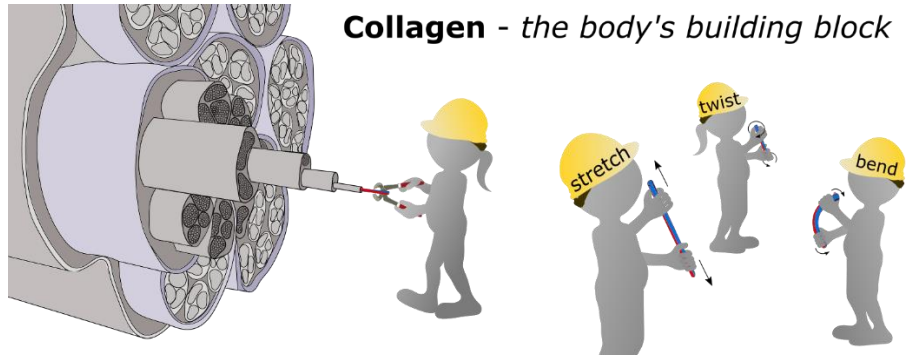

## Abstract

The primary building block of the body is collagen, which is found in the extracellular matrix and in many stress-bearing tissues such as tendon and cartilage. It provides elasticity and support to cells and tissues while influencing biological pathways including cell signaling, motility and differentiation. Collagen's unique triple helical structure is thought to impart mechanical stability. However, detailed experimental studies on its molecular mechanics have been only recently emerging. Here, we review the treatment of the triple helix as a homogeneous flexible rod, including bend (standard worm-like chain model), twist, and stretch deformations, and the assumption of backbone linearity. Additionally, we discuss protein-specific properties of the triple helix including sequence dependence, and relate single-molecule mechanics to collagen's physiological context.

## Introduction

Collagen is the most abundant protein in the human body, accounting for more than 30% of the total protein.[1,2] Collagens are defined by their unique right-handed triple helical structure, comprised of three left-handed polyproline-like helices, each with a (Gly-X-Y) repeating sequence where X and Y are often proline and hydroxyproline.[2] In the body, collagen forms ordered, hierarchical structures such as fibrils (types I, II and III) and networks (type IV).[3] The primary function of collagen is to provide elasticity, stability and support to tissues,[1] properties that contributed to its foundational role in the evolution of multicellular life.[3,4] Through its role in the extracellular matrix, collagen's mechanics are also involved in regulating cell signalling, differentiation and migration.[1,5–7] At the molecular level, collagen's flexibility imposes cellular requirements for trafficking and secretion and may regulate protein binding and hierarchical assembly.[8–13]

At the molecular level, collagen is typically viewed as a uniform triple helix, with its properties dominated by its unique quaternary structure. However, this overlooks the complexity of the native protein: the X and Y amino acid substituents vary throughout the length of each of the three constituent α-chains; the sequence varies among organisms and between collagen types (of which there are 28 in humans, localized to different tissues and consisting of homo- and heterotrimers); and there is a variety of functionally relevant post-translational modifications.[1,14–17] Collagen's size (~300 kDa; ~300 nm in length) and propensity to undergo self-association lead to challenges working with collagen *in vitro*, *in situ*, *in vivo* and *in silico*. For example, no atomistic structures of the full-length protein are available; instead, its structure has been determined using short polypeptide model systems called collagen mimetic peptides, or CMPs. The considerable difference in length between these (~10 kDa; ~10 nm) and the full-length protein is illustrated in Figure 1.[18,19] Thus, in spite of the known complexity of collagen as a protein, a reasonable initial assumption is to treat collagen as a uniform triple helix, modelled mechanically as a semiflexible rod.

Although collagen's mechanical properties are of key relevance for its physiological functions, they are surprisingly not well understood.[17] Addressing this point requires baseline information regarding the structural stability of collagen: How rigid is its signature triple helix structure? Does this rigidity vary in different chemical environments? Is this structure maintained under force? Recent years have seen the application of a variety of single-molecule approaches that aim to answer these questions (Figures 2). This review highlights these studies, most of which treat collagen as a homogeneous, flexible rod-like structure. However, collagen is not homogeneous throughout its length, and its internal structure likely influences its response.[14,20] Local structural differences shown in Figure 1 such as sequence variations,[20] mutations[18] and post-translational modifications[15] may alter the local flexibility of the collagen molecule. Thus, we also highlight recent work that begins to investigate the question of its sequence-specific mechanical stability and function.

## Collagen as a Homogeneous, Deformable Rod

When describing the mechanics of the collagen triple helix, a convenient assumption is that it behaves as a homogeneous, deformable rod, as often assumed for other helical biopolymers such as DNA and filamentous actin. The mechanics of a homogeneous rod are governed by three deformation modes – bend, stretch and twist – as illustrated in Figure 2A. The structural response of the rod to an applied force, or simply to thermal fluctuations, is determined by the elastic energy cost for each of these deformation modes, as well as potential coupling terms such as twist–stretch or twist–bend coupling.[21–23]

To extract these energy scales from experimental data, various models can be applied. The most widespread model for describing homogeneous semiflexible polymers is the worm-like chain (WLC) model.[24] In its simplest form, the standard WLC assumes an inextensible yet continuously flexible rod that is inherently linear and can undergo bending deformations: stretch and twist deformations are considered to cost considerably more energy relative to thermal energy scales and thus are disregarded when analyzing mechanical response.[25] These can be incorporated in extensions of the WLC model,[26] and we discuss them following a summary of collagen's standard WLC response.

## Bending Deformations

Collagen's bending stiffness regulates its compactness in solution (e.g. for secretory transport from the cell) and ability to conform to packing requirements of different hierarchical structures.[11,13] Bending stiffness is characterized, at a given temperature, by the *persistence length*, the length scale over which the directionality of collagen's backbone *persists*: a long persistence length correlates with a high bending stiffness (e.g. more rigid-rod-like and straight) while a shorter persistence length indicates a lower energetic cost to bend (and more flexible, compact structures). Collagen's persistence length has been evaluated using various techniques, including atomic force microscopy (AFM) imaging[17,27], optical tweezers[28–30], coarse-grained molecular dynamics (MD)[31,32], atomistic MD[33,34], electron microscopy,[35] viscometry[36] and dynamic light scattering[37] (e.g. Figure 2B). Unlike DNA, where such diverse techniques converge on a consistent value for persistence length (~50 nm),[21] estimates of collagen's persistence length vary by over an order of magnitude, from 10-170 nm.[17] Considering collagen's contour length of 300 nm, these estimates provide descriptions ranging from flexible and compact in solution to semi-rigid and quite extended. Given the physiological implications for this result, ranging from cellular transport to the material properties of collagen-based structures, it is essential to determine the cause of this discrepancy and properly describe the mechanics of this protein.

One possible source of variability is the chemical composition of collagen. Collagen can exist as homo- and hetero-trimers, and, depending on its physiological form and age, can accumulate diverse enzymatic and nonenzymatic post-translational modifications. These have been suggested to influence collagen flexibility.[16,35,38,39] Building on an early tour-de-force single-molecule study of collagen's flexibility,[35] Rezaei *et al.* recently used AFM imaging to evaluate the variability of persistence length with collagen type and source.[17] They found collagen's bending flexibility to vary little among samples, obtaining persistence lengths of ~90 nm in high-salt conditions for types I, II and III fibrillar collagens, encompassing homotrimers (II and III) and heterotrimers (I) expressed in yeast and derived from tissue.[17] This finding suggests that chemical composition does not have a strong influence on overall flexibility of collagen's triple helix.

Alternatively, chemical environment is emerging as a candidate to explain variation in persistence lengths.[17,27] Using AFM imaging, Lovelady *et al.* showed that collagen's structure was much more rigid in the presence of salt than in water.[27] In a more detailed study Rezaei *et al.* independently varied both pH and ionic strength.[17] The structures of collagen observed were strongly affected by the change in solution deposition conditions, with collagen adopting more compact configurations when deposited from solutions of lower ionic strength.[17] Persistence lengths obtained from fits to the standard WLC model appeared to decrease, from ~100 nm to ~40 nm, with decreasing ionic strength, spanning a large portion of the range of previously reported persistence lengths. However, the apparently flexible conformations at low ionic strength were not well represented by the standard WLC model, a point to which we return later in this review.

## Extensibility and Twist

Stretch- and twist-induced deformations of collagen have been less thoroughly investigated than bending. To our knowledge, there have been no direct studies of collagen's twist elasticity, with inferences about over- and underwinding of the triple helix being reached only in connection with stretch-twist coupling.[33,40,41] Extensibility of the collagen structure has been evaluated to some degree using single-molecule techniques such as optical tweezers,[42] AFM,[43] magnetic tweezers,[39–41] centrifuge force microscopy (CFM)[44] and steered MD simulations[15,45–47] (Figure 2B).

There are two general experimental approaches to establishing the stretch elasticity of collagen. The first is a more direct approach, which involves determining whether its force-extension curve is well described by the inextensible WLC model. If this model agrees with the data, and if the parameters of contour length and persistence length do not vary with the force range used for fitting, then additional deformations supplied by extensibility and twist are not needed. Most experimental collagen studies have not evaluated the force-dependence of fitting parameters and provide fits to the standard WLC up to forces of 10 pN,[28–30] the range where DNA is well described by this model.[21] In this force range, force-extension profiles provide a persistence length of 10-20 nm,[28–30,42] much shorter than the values obtained from single-molecule imaging.[17,35] Persistence lengths derived from force-extension curves may be underestimated because of the short contour length of collagen,[48] but also may not use an appropriate model for fitting: deviations from inextensible WLC behavior are suggested from preliminary force-dependent studies of fitting parameters, which find persistence length to decrease from 65 to 15 nm as the maximum force used for fitting is increased from 2 to 10 pN.[42] If validated, this finding implies a softening of the triple helix even at these low forces, resulting from additional modes of deformation such as stretch and/or twist.

The second approach uses the force dependence of enzymatic cleavage to infer changes in collagen's structure. Force-dependent enzymatic cleavage assays are based on the underlying assumption that access to a single α-chain is the rate-limiting step for collagen cleavage. Thus, if the tightness of the triple helix is destabilized by an applied force, then its cleavage rate should increase, and vice versa. Force-dependent enzymatic cleavage assays of collagen have been performed with three different proteases: two collagenases (a matrix-metalloprotease – MMP-1 – and bacterial collagenase) and trypsin. Using magnetic tweezers to stretch single collagen molecules, Adhikari *et al.* showed a force-dependent *increase* of MMP-1's cleavage rate on both a CMP[40] and full-length collagen type I.[39] They also found bacterial collagenase's cleavage rate to be force-insensitive,[39] in contrast to the result of Camp *et al.*, who also used magnetic tweezers to stretch collagen, yet found force to *reduce* its rate of cleavage by bacterial collagenase.[41] To address this controversy Kirkness *et al.* developed high-throughput single-molecule cleavage assays using centrifuge force microscopy.[44] In contrast to the studies with collagenases, they studied collagen's cleavage by trypsin, the established enzymatic probe for collagen's triple helical structure.[44,49] They found the rate of collagen's cleavage by trypsin to *increase* with force, suggesting that a stretching force destabilizes collagen's triple helical structure.[44] Of note, all of these enzymatic cleavage assays produce contradictory results below 10 pN, the range in which collagen has generally been assumed to be inextensible.

Deviations from the standard WLC model suggest further modes of mechanical deformation such as extensibility, twisting, or other force-induced structural changes of the molecule.[26,42,47] Insight into some of these potential deformations have been provided by steered MD simulations: stretching collagen by its ends results first in straightening and twisting of the triple helix, followed by helix uncoiling and breaking of hydrogen bonds (changing from a twisted "rod" to a distinct structural phase), then, as force is further increased, individual α-chain extension.[33,46,50] Hillgärtner *et al.* compared the applicability of several WLC models with AFM pulling data of collagen, finding that as collagen is stretched beyond the entropic regime, twist-stretch coupling is needed to describe the data.[26] Collagen's twist and stretch elasticity could be obtained experimentally via techniques used to uncover these properties of DNA, though the shorter contour length of collagen provides technical challenges in this regard.[22,51]

## Curved Collagen

The conventional descriptions of collagen as a rod or WLC assume that its backbone is intrinsically straight. Recent studies call this assumption into question. As mentioned above, AFM imaging studies observed compact configurations of collagen when the collagen was deposited from solutions of low ionic strength. While initially these were attributed to a short persistence length / high flexibility,[17,27] inspection of the data revealed that the standard WLC model did not agree with the length-dependent trends in flexibility.[17] Instead, a curved WLC model better described the data: conformations observed from all solution conditions could be explained by a salt- and pH-dependent curvature, with minimal variation in collagen's persistence length.[17] Because these experiments were performed on mica, they raise the question of whether collagen possesses intrinsic curvature in low-salt solutions, or whether the curvature is induced by interactions with the surface.[17] Curvature can be induced by interactions of polar, chiral filaments with a surface;[52] here, salt and pH could modulate either the strength of interactions between collagen side-chains and the mica, or the torsional stiffness of collagen, or both. Alternatively, at low salt the helical structure of collagen could possess intrinsic curvature, as seen for other coiled proteins such as tropomyosin.[53]

There is evidence of inherent curvature in collagen from other work. Collagen's structure in the native fibril possesses local curvature.[54] Also, while most MD simulations have found or assumed collagen to be intrinsically straight, some studies have found bent structures as their equilibrated CMP conformations.[45,50] Evidence for nonlinearity of CMPs has also come from a recent small angle x-ray scattering study.[55] All of these studies challenge the assumption that collagen's triple helix is intrinsically straight, demonstrating the need for further work on collagen's intrinsic shape and compactness. These features have important ramifications for the cellular energy required to package it for secretion and to bend it into a shape compatible with higher-order extracellular structure formation.

## Sequence Dependence

By treating collagen as a homogeneous rod, its protein complexity is ignored. Although homogeneous WLC analysis of collagen conformations found only minor variations with collagen source and type,[17] it is well known that sequence variability along the length of collagen results in local differences in helix pitch, dynamics, and thermal stability.[14,54,56,57] For example, distinct helical structures allow for localization of key binding proteins such as integrins, the von Willebrand Factor and matrix-metalloproteases (MMPs).[5,58–60] Additionally, a reduced triple helix stability in the MMP binding region has been proposed to facilitate collagen cleavage at that location.[12,56,57] Thermal stability varies with local sequence and particularly with imino acid content.[14,20,61] This suggests that distinct regions of the triple helix have different propensities to undergo micro-unfolding or breathing,[62,63] structural changes which may affect local bending stiffness and response to force. Dynamical structure fluctuations likely play a key role in the physiological function of collagen, particularly because the protein is thermally unstable at body temperature.[64]

To date, experimental studies of collagen's sequence-dependent bending flexibility have been limited to the pioneering work of Hofmann *et al.*[35] They determined flexibility profiles along various types of collagen imaged using electron microscopy.[35] The strongest variations in flexibility they found were along the length of type IV collagen.[35] Collagen IV is a network-forming collagen that possesses interruptions in the repetitive (Gly-X-Y) sequence; these interruptions of the triple helical structure provide increased flexibility and potential molecular recognition sites.[65]

As a predictive tool for understanding sequence-dependent flexibility of collagen, coarse-grained models show promise. Recent models have adapted techniques developed for studying DNA to parametrize collagen based on imino vs amino acid content, hydrogen-bonding between strands, and electrostatics.[66,67]

The force response of collagen has also been suggested to depend on sequence. The primary evidence for this comes from the conflicting dependence of cleavage rate on force when using MMP-1 versus bacterial collagenase, as described above.[39–41] Cleavage studies using trypsin, instead, gave results that agreed with the MMP response.[44] This finding may result from the presence of a trypsin site within the MMP region of type III collagen: MMP and trypsin could be interrogating the same region of the triple helix.[44,68] The different force response seen with bacterial collagenase could be explained by its distinct cleavage sequence and/or enzymatic mechanism.[69] How generic the response of the triple helix is to force could be investigated, for example, by use of a type III mutant in which trypsin cleavage in the MMP region is abolished.[68]

Changes in collagen's chemical composition also impact its mechanical response. Point mutations within the triple-helix region can give rise to connective-tissue diseases with mechanical phenotypes such as Osteogenesis Imperfecta and Ehlers-Danlos syndrome.[3,70] These sequence modifications can generate a local kink (packing defect) in the structure;[71] how they affect force response at the molecular level has thus far been explored only through MD simulations.[72,73] Additionally, post-translational modifications such as age-related glycation end-products (AGEs) may affect the local mechanics of collagen.[15]

## Physiologically Relevant Implications and Future Directions

Collagen's physiological function in connective tissue and extracellular matrix mechanics occurs in the context of hierarchical structures. How do its mechanics at the molecular level, as discussed in this review, impact its physiological function? Collagen mechanics continues to be extensively studied at higher levels of organization,[74] but fewer studies integrate molecular level response.[30,75–77] For example, the molecular response of collagen in strained tendons has been studied with a variety of techniques, including x-ray scattering, trypsin digestion and imaging with collagen-hybridizing peptides.[78–82] These studies have revealed that overloading the tendon results in molecular-level denaturation of collagen.[78–80] At present, it is not possible to distinguish between denaturation resulting from shear-induced extraction of an α-chain from a triple helix (perhaps via crosslinks to adjacent strained molecules[78]) and irreversible structural deformation of the triple helix (such as a change in registration of its α-chains[79]). Connecting single-molecule mechanics with higher-order mechanical response is a key challenge and an area for growth in collagen research.

At the molecular level, much work remains to be done to understand the mechanics of collagen, and how they are influenced by the local sequence. Most of this review focused on describing collagen in the context of the widely-used inextensible WLC model. However, evidence is accumulating that deformations besides bending, such as twist and stretch, are accessible at low forces and perhaps even contribute noticeably in unloaded collagen. It is likely that couplings between these three deformation modes are also of importance. It is of interest to compare collagen and DNA mechanics as much has been learned over the preceding decades about the energy scales governing DNA's bend, twist and stretch dynamics. These deformations of DNA enable its compaction into viruses, around nucleosomes, and manipulation by many regulatory proteins. Similarly, understanding the energy scales required to bend, stretch and twist collagen will provide desperately needed information about its ability to be trafficked in constrained geometries and its capacity to regulate its binding and manipulation by regulatory partners. We hope that this brief review stimulates many more studies of collagen, a protein with intrinsically mechanical roles that possesses many mysteries yet to be unravelled.

Due to limitations of the article size, we recognize that this article is not completely comprehensive and apologize for any omissions.

## Acknowledgements

The authors acknowledge funding from the Natural Sciences and Engineering Council of Canada (NSERC) (Discovery Grant to NRF) and from the Deutsche Forschungsgemeinschaft (DFG) (postdoctoral fellowship to KL). We thank many past and current members and collaborators of the Forde lab for their critical reading of a draft of this manuscript.

## Conflict of Interest

MWHK is employed as a consultant for 3Helix.

Possible experimental reasons for the wide variation in reported persistence lengths for collagen were investigated. Using atomic force microscopy imaging, the authors found a persistence length of ~90 nm, which did not depend strongly on collagen type (I, II and III) or source (recombinant vs tissue-derived). Surprisingly, they found that collagen possessed (either intrinsic or surface-induced) curvature at low salt concentrations.

The standard WLC model and multiple extensions of it were described and applied to a previously reported force-extension measurement of collagen. Estimates of persistence lengths and stretch and twist moduli were provided from testing the applicability of each model, though these are based on very limited experimental data.

The first single-molecule stretching experiment of collagen using optical tweezers. The reported persistence length of ~15 nm suggested collagen to be a highly flexible polymer.

A seminal study that was the first to apply statistical interpretations of chain conformations to analyse the "sequence-dependent" flexibility of a variety of collagen types imaged using rotary shadowing electron microscopy (though the sequences were not all known at the time).

Magnetic tweezer assays were performed to directly compare collagen's force-dependent cleavage rate by MMP-1 and bacterial collagenase, which had provided conflicting results in earlier studies using different substrates.[40,41] Here, collagen's cleavage by MMP-1 was shown to be accelerated with force while cleavage by bacterial collagenase was not.

The first application of centrifuge force microscopy to study enzymatic processes. The authors applied the widely used trypsin cleavage assay for collagen stability and found that the triple helix was destabilized when stretched by a force of ~10 pN.

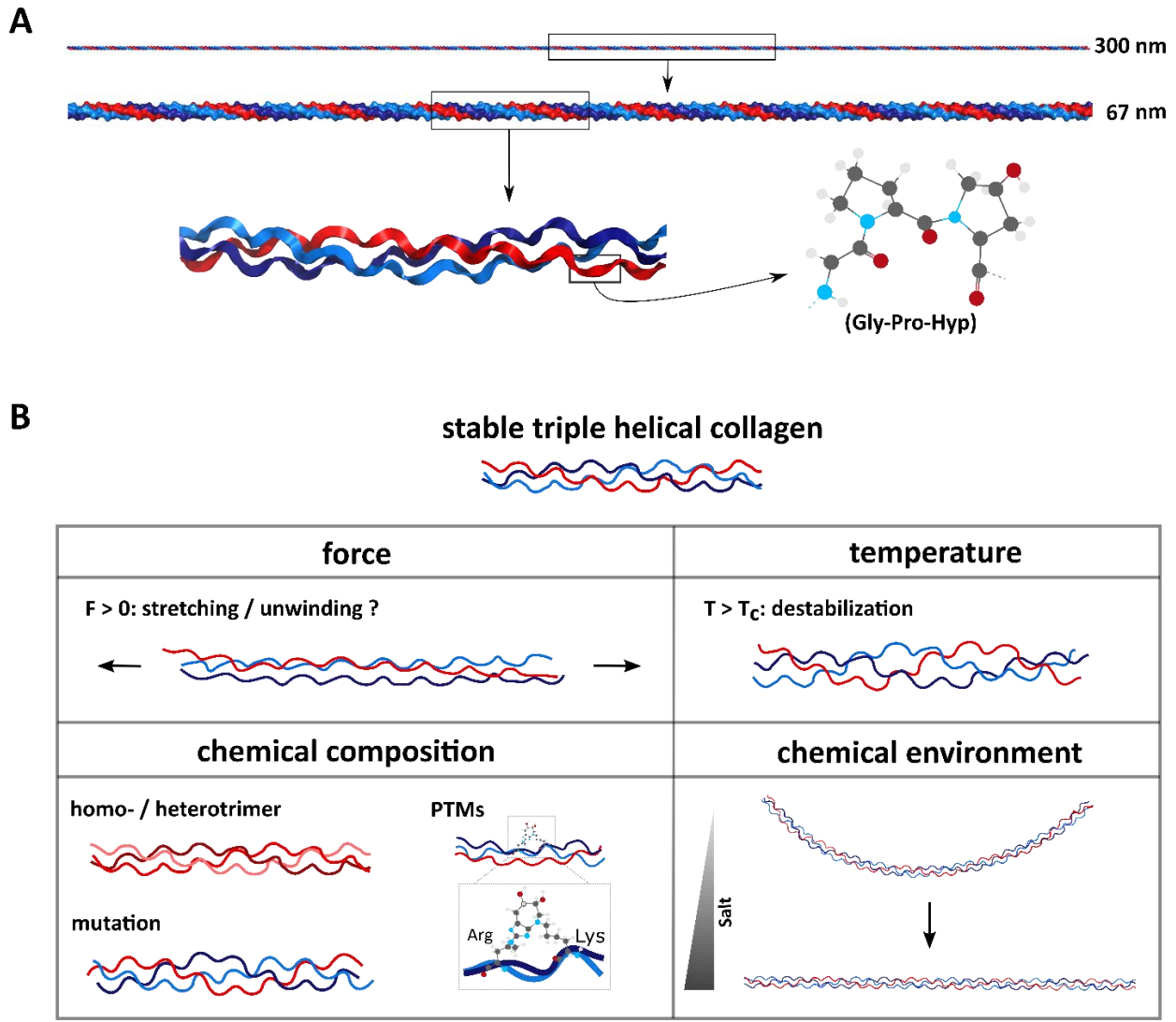

*Figure 1 Schematic representation of collagen's structure and the influence of internal and external factors on its structure.* A) Structural depictions of the most-studied length scales of collagen. From top: Full-length type I collagen structure (300 nm) as predicted by The BuScr collagen-building script,[61] in which the three α-chains are represented in dark blue, light blue and red. One fibril-repeat length of this structure (67 nm) that includes the MMP binding region and exhibits structural heterogeneity (differences in helical pitch). Crystal structure of a collagen-mimetic peptide (CMP) (~10 nm), which includes the MMP-1 binding site (PDB: 4AUO).[83] Atomic representation of the tripeptide sequence glycine-proline-hydroxyproline; glycine residues are obligatory every third amino acid in collagen's sequence. B) Influence of force, temperature, chemical composition and chemical environment on the triple helical structure. Schematics adapted from the work of others are from PDB 1CAG[18] (mutation), from reference [15] (age-related post-translational modification, PTM), from reference [50] (force) and from reference [17] (chemical environment).

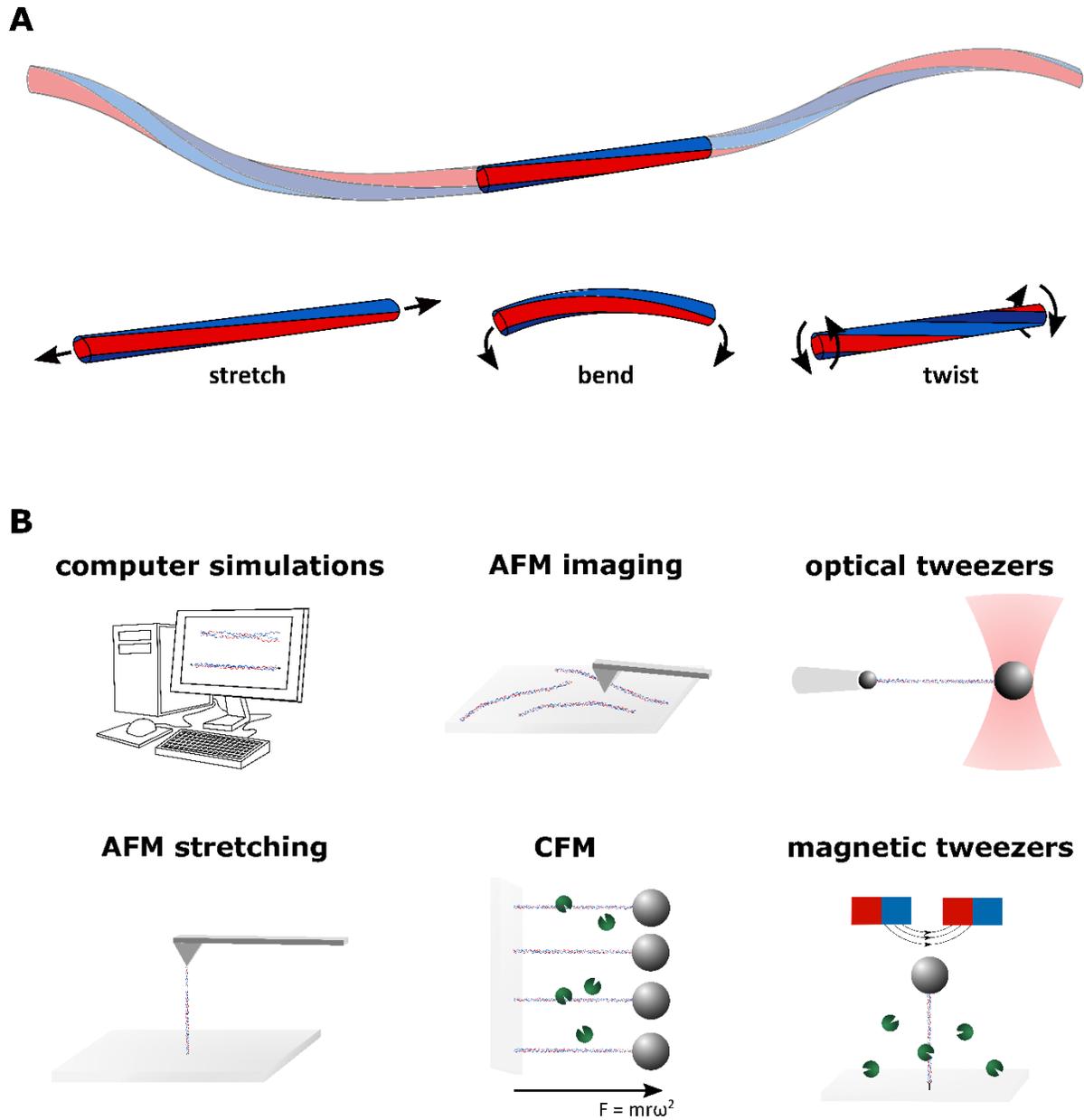

*Figure 2 Biophysical analysis of collagen's mechanical properties.* A) Mechanics of a homogeneous rod can be described by three elastic energies: stretch, bend, and twist. B) Illustration of techniques that have been used to study collagen's mechanical properties at the single-molecule level (not to scale).